\documentclass[times,5p,twocolumn]{elsarticle}

\journal{arXiv}

\usepackage{amsmath,amsfonts,amssymb,amsthm} 
\usepackage{graphicx} 
\usepackage[export]{adjustbox} 
\usepackage{bigstrut} \usepackage{multirow} \usepackage{rotating}
\usepackage{hyperref}
\usepackage{colortbl}
\usepackage{algorithm}
\usepackage{algpseudocode}

\def\real{\mathbb{R}}
\def\setZ{\mathbb{Z}}

\DeclareMathOperator*{\argmin}{arg\,min}
\newcommand{\set}[1]{\left\{#1\right\}}
\newcommand{\setn}{\{1,2,\dots,n\}}
\newcommand{\sett}[1]{\{1,2,\dots,#1\}}
\newcommand{\seq}[1]{\left\langle#1\right\rangle}
\newcommand{\floor}[1]{\left\lfloor #1 \right\rfloor}%
\newcommand{\ceil}[1]{\left\lceil #1 \right\rceil}%
\newcommand{\norm}[1]{\left\lVert#1\right\rVert}
\newcommand{\mat}[1]{\begin{bmatrix}#1\end{bmatrix}}
\newcommand{\equ}[1]{\begin{equation}#1\end{equation}}
\newcommand{\lequ}[2]{\begin{equation}#2\label{eq:#1}\end{equation}} 
\newtheorem{example}{Example}[section]

\begin{document}
\begin{frontmatter}
\title{Closed form distance formula for the balanced multiple travelling salesmen}

\author{Wolfgang Garn\fnref{w.garn@surrey.ac.uk}}
\address{Department of Business Transformation, University of Surrey, United Kingdom}


\begin{abstract}
As a first contribution the mTSP is solved using an exact method and two heuristics, 
where the number of nodes per route is balanced.
The first heuristic uses a nearest node approach
and the second assigns the closest vehicle (salesman).
A comparison of heuristics 
with test-instances being in the Euclidean plane showed
similar solution quality and runtime.
On average, the nearest node solutions are approximately one percent better.
The closest vehicle heuristic is especially important when the nodes (customers) are not known in advance, e.g. for online routing.
Whilst the nearest node is preferable when one vehicle has to be used multiple times to service all customers.

The second contribution is a closed form formula that describes the mTSP distance
dependent on the number of vehicles and customers.
Increasing the number of salesman results in an approximately linear distance growth
for uniformly distributed nodes in a Euclidean grid plane.
The distance growth is almost proportional to the square root of number of customers (nodes).
These two insights are combined in a single formula.
The minimum distance of a node to $n$ uniformly distributed random (real and integer) points 
was derived and expressed as functional relationship dependent on the number of vehicles.
This gives theoretical underpinnings and is in agreement with the distances found via the previous mTSP heuristics.
Hence, this allows to compute all expected mTSP distances without the need of running the heuristics.
\end{abstract}


\end{frontmatter}

\section{Introduction}\label{sec:Introduction} 

The multiple Travelling Salesman Problem (mTSP) 
has $m$ salesmen (vehicles)  which have to visit $n$ customers (nodes).
More precisely a customer will be visited exactly once by one and only one salesman.
The salesmen are located in  a single depot (source node).
Salesmen have to return to their starting point. 
One could emphasise this by stating the salesmen travel closed loops.
If they do not have to come back to the source node, 
then this is known as an \textit{open} mTSP.

The mTSP is a special case of the Capacitated Vehicle Routing Problem (CVRP).
The difference is that customers have a demand and vehicles (e.g trucks) have a capacity limit.
Hence, VRP algorithms can be used for the mTSP as well.
On the other hand the TSP is a special case of the mTSP with $m=1$.
There are various different flavours of the mTSP 
differentiated by means of objective functions and constraints.
In this paper we will focus on minimising the total distance
subject to an upper and lower limit of customers a salesman has to visit.
These bounds are used for node balancing.
Other constraints (not considered in this study) deal with multiple depots and time windows.

Several solution approaches have been investigated, 
which can be roughly grouped into:
\begin{itemize}
	\item Greedy heuristics,
	\item Local search algorithms,
	\item Meta heuristics, and
	\item Exact methods.
\end{itemize}
Usually the runtime of these approaches increase is in the order they were listed. 
The mTSP is an NP-Complete problem \cite{lenstra1981complexity}.
Classic meta heuristics to solve the mTSP include 
genetic algorithms (GA) \cite{tang2000multiple, brown2007grouping, yuan2013new}, 
simulated annealing (SA) implementations \cite{paydar2010applying, song2003extended, turker2016gpu,liu2018research}
and tabu search (TS) approaches \cite{golden1997adaptive, ryan1998reactive, song2003extended}.
Interestingly, GA are more often used than SA and TS when solving the mTSP.
Recently, \citet{nazari2018reinforcement} used neural networks with reinforcement learning 
to solve mTSP instances with similar structure. mTSP instances were apparently solved efficiently, after an initial resource consuming learning phase.
Neural networks were applied to solve the mTSP far earlier. 
In 1989 \citet{wacholder1989neural} used Neutral Networks based on a
Hopfield-Tank's neuromorphic city-position map using the Basic Differential Multiplier Method which evaluates Lagrange multipliers simultaneously.
Later \citet{hsu1991study} used a neural network based on a self-organised feature map model to solve the mTSP.
Another interesting neural network based on competitions was done by \citet{somhom1999competition}.
Exact methods for the mTSP are based on integer programming formulations.
\citet{bektas2006multiple} gives an overview and several solution procedures.
Typically they are solved with brand and bound \cite{ali1986asymmetric,gromicho1992exact},
or in conjunction with Lagrangian relaxation \cite{gavish1986optimal}.
Cutting planes are another popular approach which is introduced by \citet{laporte1980cutting} to solve the mTSP.

The aim of this work is to gain insights into the mTSP dynamics 
when the number of vehicles changes.
Two greedy heuristics are introduced (Section \ref{sec:Heuristics}). 
They are compared against each other and an exact method (Section \ref{sec:EmpiricalInsights}). 
We compare these heuristics to implementations by other authors as well.
A functional relationship 
describing the solution distance depending on 
the number of travelling salesmen and customers is given
for uniformly distributed customer locations (Section \ref{sec:FunctionalRelationship}).
The first steps of an exact approach to proof this function are presented.

\section{Problem Formulation}\label{sec:ProbFormulation}
The mTSP is defined on a network 
with nodes $N = \setn$, arcs $A = \seq{a_1,\dots,a_m}$ and
distances $d = \seq{d_{a_1},\dots, d_{a_m}}$.
Instances in the Euclidean plane often have a complete underlying graph structure.
In this case it is more convenient to use a distance matrix $D = (d_{ij})$.

We will base the exact formulation on \citet{kara2006integer}.

The objective function is:
\lequ{obj}{f(x)=\sum_{(i, j) \in A} d_{i j} x_{i j}.}
Here, $x_{i j}$ is the binary decision variable 
whether node $j$ is visited coming from node $i$.

This is subject to several constraints.
We need to ensure $m$ salesmen leave and enter depot, which is located in node 1:
\lequ{c.depot}{\begin{array}{l}{\sum_{j=2}^{n} x_{1 j}=m} \\ {\sum_{j=2}^{n} x_{j 1}=m}\end{array}.}

On each route exactly one node with the exception of the depot is entered and left by one vehicle:
\lequ{c.visit}{\begin{array}{ll}{\sum_{i=1}^{n} x_{i j}=1,} & {j=2, \ldots, n} \\ {\sum_{j=1}^{n} x_{i j}=1,} & {i=2, \ldots, n}\end{array}}

Now there is a minimum number $K$ and maximum number $L$ of nodes 
that have to be visited by each salesman.
The decision variables $u_i \in \setZ (i \in N)$ 
keep track of the sequence and eliminate subtours.
Each salesman tour has an upper bound:
\lequ{c.upper bound}{u_{i}+(L-2) x_{1 i}-x_{i 1} \leqslant L-1, \quad i=2, \ldots, n;}
and a lower bound:
\lequ{c.lower bound}{u_{i}+x_{1 i}+(2-K) x_{i 1} \geqslant 2, \quad i=2, \ldots, n.}

Furthermore, we will not allow short tours, i.e. $K>2$ or:
\lequ{c.no return}{x_{1 i}+x_{i 1} \leqslant 1, \quad i=2, \ldots, n.}

The actual subtour elimination constraints are:
\lequ{c.st elim}{u_{i}-u_{j}+L x_{i j}+(L-2) x_{j i} \leqslant L-1, \quad 2 \leqslant i \neq j \leqslant n.}

Existing exact algorithms find solutions 
for small instances (number of nodes less than 20) in seconds, 
for medium instances (number of nodes 20 to 100) in hours,
and for large instances they generally ``struggle''.
Hence, it is important to use heuristics.

\section{Heuristics}\label{sec:Heuristics}
We will consider two greedy heuristics. 
The first heuristic uses a nearest node approach, 
and the second one assigns the closest vehicle (salesman). 
The input for the algorithms is a distance matrix and the number of salesmen $m$.
The output is a list of $m$-routes.

The nearest node approach builds a route for each vehicle iteratively.
In the vehicle 's iteration the route is initialised with the source node ($f=1$).
Within the iteration a loop adds the nearest node to the last visited node
using the nodes that have not been visited yet.
Once the route reaches a certain number of nodes the next vehicle's iteration starts.

Algorithm \ref{alg:mTSP-nearest-node} shows the details.
A distance matrix $D$ and the number of salesmen $m$ are required as input.
The elements in the distance matrix's diagonal  must be $\infty$.
\begin{algorithm}[!h]
    \caption{Nearest Node mTSP. \label{alg:mTSP-nearest-node}}
    \begin{algorithmic}[1]
    \Require distance matrix $D=(d_{ij}) \in \real^{n \times n}$; number of salesman $m$; 
    \Ensure routes $r = \seq{r_1,\dots,r_m}$

		\State $v=\set{2,\dots,n}$ \Comment{nodes not visited}
		\For { $k$ in \sett{$m$} }
      \State $r_k=1$; $f=1$; \Comment{init route and from node}
			\While{$\#v>0$ and $\#r_k\leq M$}
				\State $j = \argmin {D}_{f,v}$; 
        $t = v_j$; 
        \State ${D}_{v,t} = \infty$; \Comment{ensure no one goes to $t$}
        \State $r_k = r_k \cup t$; $v = v \setminus t$; \Comment{add $t$ to route and remove from not visited nodes}
        \State $f = t$; \Comment{update from node}
			\EndWhile
		\EndFor
    \end{algorithmic}
\end{algorithm}
The algorithm returns a sequence of balanced routes.
The routes are balanced by having 
\lequ{trucks_a}{a = n-1-m \floor{\frac{n-1}{m}}}
salesmen, which visit $M_u =\ceil{\frac{n-1}{m}}$ customers (not counting the source node).
Otherwise, there are  $M_l=\floor{\frac{n-1}{m}}$ or less customers.
This allows us to define a conditional $M$:
\lequ{customers_per_truck}{M = [k\leq a]M_u + [k>a]M_l,}
where $k$ is the salesman index number.
The salesman number is set in line 2.
Line 3 initialises the route sequence 
with source node 1.
Line 4 starts a loop in case there are nodes that have not been visited yet,
and the salesman's customer limit $M$ has not been exceeded.
The index $j$ and node $t$ the salesman will visit next are determined in line 5 
by taking row vector $f$ from $D$ and considering $v$ (the nodes not visited yet).
For subsequent steps arcs towards $t$ are unnecessary.
Hence, respective distances are set to infinity (line 6).
Line 7 adds $t$ to the current route 
and removes it from the not visited node list.
The last visited node $t$ becomes the starting point for the next search (line 8).
The while-loop completes route $k$. 
The for-loop iteratively creates all routes $r = \seq{r_1,\dots,r_m}$.
\begin{example}[Two routes]
Assume the coordinates $C$ for nine nodes ($N = \sett{9}, n=9$) are given:
$$C = \mat{10 & 7 & 2 & 3 & 6 & 12 & 16 & 19 & 14 \\
           5 & 8 & 7 & 3 & 2 & 6 & 8 & 4 & 1 }.$$
The distance matrix $D$ is obtained by computing Euclidean distances:
$$ d_{ij} = \norm{C_i - C_j}_2 = \sqrt{(C_{i1}-C_{j1})^2 + (C_{i2}-C_{j2})^2}, $$
where $i$ and $j$ are nodes having coordinates $C_i$ and $C_j$.

Node 1 represents the source. Two salesman $m=2$ are available.
According to Equation \ref{eq:customers_per_truck} a salesman will visit $\frac{n-1}{m}=4$ customers (not counting the source node).
Figure \ref{fig:mTSP-example-solution} shows the optimal solution,
\begin{figure}[htbp]\center
	\includegraphics[width=\columnwidth, height=4cm, keepaspectratio]{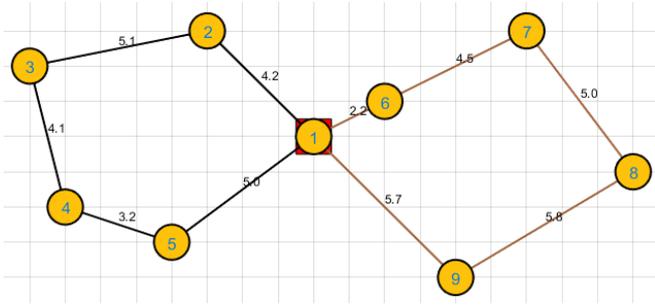}%
	\caption{mTSP example solution}%
	\label{fig:mTSP-example-solution}%
\end{figure}
which was found by using Algorithm \ref{alg:mTSP-nearest-node}. 
However, in general, the algorithm cannot guarantee finding the optimal solution.
Looking at the solution it can be observed that travelling salesman 1 first visits node 6, because it is the nearest node from 1.
The nearest node (not visited yet) from 6 is 7 with a distance of 4.6.
The algorithm continues that way and forms route $r_1=\seq{1, 6, 7, 8, 9}$.
Since it is a closed tour node 1 is added to the route sequence.
This route's distance is $\sim$ 23.2.
Route 2 is found in a similar way having distance $\sim$ 21.6.
Hence, the total distance is $\sim$ 44.8.
\end{example}

This algorithm naturally arises if there is only one
salesman (truck, vehicle, forklift, car) available and the capacity limit is $M$ 
leading to $m$ tours a single salesman has to undertake.

If a fleet of $m$ trucks is available, i.e. there are $m$ salesman
the closest vehicle approach is better.
The idea is to assign the salesman, which is closest to the not yet visited nodes.
Algorithm \ref{alg:mTSP-closest-vehicle} shows the details for this implementation.
\begin{algorithm}[!h]
    \caption{Closest Vehicle mTSP. \label{alg:mTSP-closest-vehicle}}
    \begin{algorithmic}[1]
    \Require distance matrix $D=(d_{ij}) \in \real^{n \times n}$; number of vehicles $m$; 
    \Ensure routes $r = \seq{r_1,\dots,r_m}$
		\State $f=e_m=\mat{1&1&\dots&1}$ \Comment{all vehicle start from node 1} 
		\State $r=e_m$; \Comment{initialise routes}
		\State $v=\set{2,\dots,n}$; \Comment{nodes not visited}
		\State $y = \mat{0,0,\dots,0}$; \Comment{number of nodes visited per vehicle}
		\While { $\#v > 0$}
		  \State $e = \min_c{{D}_{f,v}}$; \Comment{min for each column}
			\State $s = \argmin_c{{D}_{f,v}}$; \Comment{vehicles}
			\State $k = \argmin e$; \Comment{identify vehicle index}   
			\State $t = v_k$; \Comment{to node}
      \State $r_k = r_k \cup t$; \Comment{add $t$ to route}
			\State ${D}_{v,t} = \infty$; \Comment{ensure no one goes to $t$}
			\If{$k \leq a$}
				\If{$y_k \geq \ceil{\frac{n-1}{m}}$}
					\State $D(t, v) = \infty$; \Comment{cannot go beyond this node}
				\EndIf
			\Else
				\If{$y_k \geq \floor{\frac{n-1}{m}}$}
					\State $D(t, v) = \infty$; \Comment{cannot go beyond this node}
				\EndIf
			\EndIf
			\State $v = v \setminus t$; 
			\State $f_k = t$; \Comment{new from node}
		\EndWhile
    \end{algorithmic}
\end{algorithm}
Line 1 and 2 ensure that the vehicles start from node 1.
Line 3 states that all nodes except node 1 have to be visited.
$y$ in line 4 will keep track of how many nodes a vehicle has visited.
In line 5 a loop is started that will only stop once all nodes have been visited.
The minimum distance from each vehicle to the available nodes is computed (line 6).
Line 7 gives the corresponding index, which is equivalent to a vector of vehicles.
The index of the minimum distance identifies the required vehicle index $k$ (line 8).
Finally, the ``to-node'' is obtained (line 9).
We add this node to the appropriate route (line 10).
Ensure that no vehicle will go to the found node anymore (line 11).
Line 12-20 check whether a vehicle has exhausted its capacity.
In case a salesman's limit has been exceeded the remaining distances are set to infinity.
This prevents the salesman to be chosen for any further routing.
$t$ was the last node on vehicle $k$'s route.
Line 21 removes $t$ from the not-visited-nodes list.
Line 22 specifies vehicle $k$'s last visited node $f_k$.
Vector $f$ is the origin for subsequent routes.

This algorithm occurs in applications, where a fleet is available.
Here, each vehicle ``bids'' for the next job.
The vehicle with the shortest distance to the next job is allocated to it.
As mentioned in the acknowledgement a similar scenario was the motivation for developing the closest vehicle heuristic.
The actual scenario consisted out of a fleet of autonomous forklift-trucks (robots equipped with Artificial Intelligence),
which operate in a warehouse. They move pallets between storage locations or ports.
However, pallet movements are not known in advance. 
Hence, the closest vehicle heuristic assigns the forklift-truck nearest to the storage location to the next job.
The subsequent section gives insights about the performance of the heuristics.

\section{Empirical Insights}\label{sec:EmpiricalInsights}
The exact algorithm (A), nearest node heuristic (B) and the closest vehicle algorithm (C) 
are compared with each other in regard to distances (see Table \ref{tab:mTSP-small-test}).
\begin{table*}[htbp]
  \centering
  \caption{mTSP distance comparison for small test instances.}
	\adjustbox{width=.9\textwidth}{%
    \begin{tabular}{rrrr|rrr|rrr}
    \hline
     \#nodes & $m$ & K & L &  dist A &  dist B &  dist C &  dist\% AB & dist\% AC &  dist\% BC \bigstrut\\
    \hline
    garn9 & 2 & 2 & 5 & 44.8 & 44.8 & 44.8 & 0 & 0 & 0 \bigstrut[t]\\
    garn13 & 3 & 3 & 4 &  57.7 & 64.7 & 69.7 & 12.1 & 20.8 & 7.7 \\
    garn13 & 3 & 3 & 6 & 48.9 & 64.7 & 69.7 & 32.2 & 42.4 & 7.7 \\
    garn20 & 3 & 3 & 7 & 90.5 & 97 & 104.7 & 7.2 & 15.6 & 7.9 \\
    bays29 & 4 & 4 & 8 & 2603 & 3482 & 3482 & 33.8 & 33.8 & 0 \\
		eil51 & 2 & 24 & 25 & 444.1 & 533.9 & 619.8 & 20.2 & 39.6 & 16.1 \\
    berlin52 & 5 & 10 & 11 & 9845.2 & 12027.4 & 12593.4 & 22.2 & 27.9 & 4.7 \bigstrut[b]\\
    \hline
    \end{tabular}%
		} 
  \label{tab:mTSP-small-test}%
\end{table*}%
A set of six small test instances with node numbers varying between 9 and 51 nodes are used. 
These test instances can be downloaded from \url{wiki.smartana.org}.
The instances are further refined by providing the number of salesmen ($m$),
the minimum $K$ and maximum number $L$ of customers visited.
However, for the heuristics replace $K$ and $L$ with $M$ using Equation \ref{eq:customers_per_truck}.
The exact mTSP algorithm makes use of the initial solution from the nearest node heuristic. 
By the way, this reduced the algorithm's runtime by about 20\%.
Considering the test instance with 29 nodes - the exact algorithm is 33.8\% better than the heuristics. 
The greedy heuristic is 4.7\% better on average than the closest vehicle heuristic. 
The heuristics return solutions almost instantly.
The exact method's runtime is a few seconds for the instances up to 29 nodes 
on a computer with 16GB RAM and processor Intel(R) Core(TM) i7-4600 CPU \@ 2.10 GHz 2.70 GHz.
The instances with 51 and 52 nodes run for several minutes before returning the optimal solution.

Table \ref{tab:table-classic-medium-instances} uses ``classic'' TSP test instances
originating from the TSP library.
These instances are often used for the mTSP (\cite{zhou2018comparative, venkatesh2015two}).
\begin{table*}[htbp]
  \centering
  \caption{mTSP distance comparison for classic medium test instances.}
\begin{tabular}{rl|rrrr}
\hline
\multicolumn{1}{l}{Instance} & Algorithm & \multicolumn{1}{l}{$m=2$} & \multicolumn{1}{l}{$m=3$} & \multicolumn{1}{l}{$m=4$} & \multicolumn{1}{l}{$m=5$} \bigstrut\\
\hline
\multicolumn{1}{l}{eil51} & nearest &        533.91  &        613.42  &        640.88  &        685.97  \bigstrut[t]\\
  & closest &        619.78  &        634.93  &        669.88  &        701.98  \\
  & exact &        444.09  &        464.11  &        499.76  &        535.98  \bigstrut[b]\\
\hline
\multicolumn{1}{l}{eil76} & nearest &        692.58  &        747.34  &        782.76  &        874.24  \bigstrut[t]\\
  & closest &        692.58  &        804.21  &        841.31  &        839.86  \\
  & exact &        558.59  &        589.11  &        653.88  &        709.18  \bigstrut[b]\\
\hline
\multicolumn{1}{l}{eil101} & nearest &        789.40  &        846.99  &        872.21  &        967.82  \bigstrut[t]\\
  & closest &        887.81  &        899.68  &        913.17  &        990.98  \\
  & exact &        698.72  &        722.53  &        859.94  &        879.89  \bigstrut[b]\\
\hline
\multicolumn{1}{l}{kroA100} & nearest &   30,180.00  &   32,064.00  &   32,988.00  &   36,337.00  \bigstrut[t]\\
  & closest &   30,180.00  &   32,967.00  &   34,520.00  &   36,998.00  \\
  & exact &   24,649.54  &   29,476.83  &   32,988.00  &   36,336.53  \bigstrut[b]\\
\hline
\multicolumn{1}{l}{kroA150} & nearest &   35,881.00  &   36,411.00  &   39,657.00  &   41,786.00  \bigstrut[t]\\
  & closest &   35,633.00  &   38,439.00  &   42,074.00  &   46,503.00  \\
  & exact &   34,448.19  &   36,411.12  &   39,656.53  &   41,785.88  \bigstrut[b]\\
\hline
\multicolumn{1}{l}{kroA200} & nearest &   36,383.00  &   42,517.00  &   45,157.00  &   46,631.00  \bigstrut[t]\\
  & closest &   40,313.00  &   42,268.00  &   44,808.00  &   45,383.00  \\
  & exact &   36,382.81  &   42,517.00  &   45,156.89  &   46,631.00  \bigstrut[b]\\
\hline
\end{tabular}%

  \label{tab:table-classic-medium-instances}%
\end{table*}%

Six instances with nodes varying between 51 and 200 were used.
The number of salesmen were varied between 2 and 5.
The first node was used as depot to which all salesmen had to return.

Three implementations are compared against each other. 
Again, the nearest neighbour, closest vehicle heuristic and exact method are used.
The formulations and algorithms can be found in the previous sections.
I used the nearest neighbour heuristic as start solution for the exact method.
The computational runtime was restricted to eight hours (Note: it can take over an hour for the problem formulation).
The maximum number of nodes with a route was restricted to: $L = \ceil{\frac{n-1}{m}}$.
The minimum was $K = L-1$.
The optimal solution was found for eil51 with $m \in \set{2,3}$ and eil76 for $m=2$.
The problem formulation plus computational time for these instances were 12.6, 62.6 and 55.5 minutes.
It can be seen that all eil* test instances achieved a solution improvement.
Only a few kro* instances achieved a solution improvement from the nearest neighbour heuristic using the exact method.
In reflection $L$ and $K$ were too restrictive.
The computational performance may be improved by using the initial solutions as constraints.
For instance, for each route $r_k$ we can require:
\equ{\sum_{(i,j)\in r_k}{d_{ij}x_{ij}} \leq \max_{k \in \sett{m}}h_k, ~k\in \sett{m},}
where $h_k$ is the maximum distance a salesman travelled using the nearest neighbour heuristic.
Furthermore, the total distance can be limited as well: $ f(x) \leq \sum_{k=1}^m {h_k}$.
This will help to cut several branches within the branch and bound algorithm.
An easier approach is to increase the runtime.

\cite{zhou2018comparative, venkatesh2015two}) compared other mTSP heuristics.
\citet{zhou2018comparative} have experimental studies, which discuss the solution improvement and the required runtime. 
For the mTSP a table with six test instances (between 52 and 200 nodes) is given 
that compares four heuristics against each other.
Three of them were from the authors 
and the \textit{invasive weed optimisation} algorithm (IWO) was introduced by \citet{venkatesh2015two}.
Their mTSP implementation uses multiple depots 
located indirectly via the requirement of being on the salesman's route.
The minimum number nodes visited can be set. 
In their experiments it was set to one.
A direct comparison of the results in this study with theirs is only possible to a certain extend. 
This is due to their problem design being more relaxed, which means it should lead to better solutions.
Furthermore, their algorithms are meta-heuristics, which is usually another reason better solutions.
Yet, it is interesting to note that the results reported by \citet[p572, Table 4]{zhou2018comparative}
are of similar magnitude as the heuristics introduced, here, for the $m=2$ and instances up to 150 nodes.
Even more surprising is that the reported kro* instances $n>100$ and $m>2$ distances exceed significantly
the distances reported here; despite using a meta-heuristic.
\citet{zhou2018comparative} determine multiple routes without the constraint
to come back to a single depot. That means, all distances should have been lower.
Hence, the heuristics here with slight adaptations should outperform the
genetic algorithms (GAs), 
which means that the nearest node and closest vehicle heuristics should be used to find initial solutions for the GAs.

\section{Functional Relationship}\label{sec:FunctionalRelationship}
Similar to the previous section test instances are computed and compared.
However, in this section we will give the functional relationship
that relates the solution distance to the number of salesmen and customers (nodes).
The coordinates for $n$ nodes are uniformly distributed pseudo-random integers in $X^2$ where $X=\sett{100}$.
They are generated for test-instances ranging from 50 to 500 nodes (step size 50).
The Euclidean distance matrix is computed from these coordinates.
Hence, the number of arcs is $n^2-n$ leading up to 249,500 arcs.
Since, the matrix is symmetric the number of arcs could be reduced, 
and the subsequent algorithms could be adjusted for efficiency reasons.
However, in a real-world environment symmetry cannot be guaranteed.
For each test-instance category 30 samples are generated.
The nearest node and closest vehicle algorithms are compared against each other.
The number of travelling salesmen $m$ is varied between 2 and 7.

\begin{figure*}[htbp]
	\centering
			\begin{tabular}{cc}
					\includegraphics[width=0.45\textwidth]{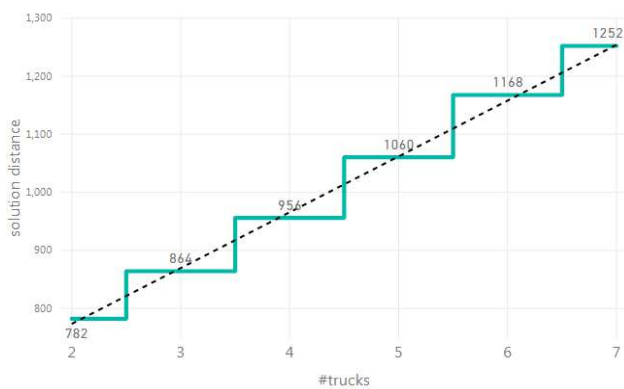} &
					\includegraphics[width=0.45\textwidth]{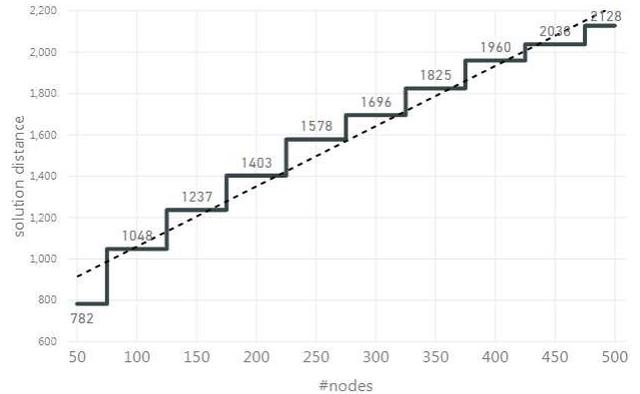} \\
					(a) & (b)
			\end{tabular}
	\caption{Average solution distance for the closest vehicle heuristic in relation to 
	         (a) number of trucks for 50 nodes;
					 (b) test instance size (\#nodes) for $m=2$.}
	\label{fig:dist-relation}
\end{figure*}

Table \ref{tab:heuristics-medium-instances} shows the averages distances $s_{tm}$ and their standard deviations $\sigma_{tm}$.
$t$ is the number of nodes and $m$ is the number of salesmen.
\begin{table*}[htbp]
  \centering   
	\caption{Average solution distances $\pm$ standard deviations for heuristics 
		for medium test-instances with varying number of salesmen $m$.}
\adjustbox{width=\textwidth}{
    \begin{tabular}{rl|cccccc}
    \hline
		\#nodes & Algorithm & $m=2$ & $m=3$ & $m=4$& $m=5$& $m=6$& $m=7$\bigstrut\\
    \hline
    50 & nearest node & 772.2$\pm$51.0 & 854.3$\pm$64.4 & 957.2$\pm$82.7 & 1,044.5$\pm$76.8 & 1,155.6$\pm$99.5 & 1,241.3$\pm$104.4 \bigstrut[t]\\
    50 & closest vehicle & 782.3$\pm$53.3 & 864.0$\pm$80.3 & 956.1$\pm$79.7 & 1,060.5$\pm$80.7 & 1,167.6$\pm$94.6 & 1,252.2$\pm$102.4 \bigstrut[b]\\
    \hline
    100 & nearest node & 1,033.5$\pm$69.9 & 1,112.6$\pm$61.9 & 1,203.5$\pm$76.8 & 1,266.9$\pm$72.5 & 1,371.5$\pm$86.1 & 1,457.4$\pm$108.7 \bigstrut[t]\\
    100 & closest vehicle & 1,048.1$\pm$64.2 & 1,109.9$\pm$52.2 & 1,228.6$\pm$80.1 & 1,305.8$\pm$79.9 & 1,384.4$\pm$78.4 & 1,495.9$\pm$91.3 \bigstrut[b]\\
    \hline
    150 & nearest node & 1,231.9$\pm$59.3 & 1,305.2$\pm$52.2 & 1,360.6$\pm$53.5 & 1,464.9$\pm$69.2 & 1,550.4$\pm$72.7 & 1,647.4$\pm$82.7 \bigstrut[t]\\
    150 & closest vehicle & 1,236.6$\pm$61.3 & 1,322.8$\pm$61.5 & 1,406.6$\pm$64.5 & 1,503.4$\pm$80.7 & 1,567.5$\pm$80.0 & 1,657.9$\pm$77.9 \bigstrut[b]\\
    \hline
    200 & nearest node & 1,419.9$\pm$68.1 & 1,519.8$\pm$70.6 & 1,577.4$\pm$92.1 & 1,679.8$\pm$80.8 & 1,776.3$\pm$97.6 & 1,864.3$\pm$124.6 \bigstrut[t]\\
    200 & closest vehicle & 1,403.0$\pm$63.9 & 1,518.1$\pm$60.9 & 1,582.5$\pm$86.9 & 1,693.8$\pm$72.0 & 1,779.6$\pm$100.8 & 1,877.8$\pm$124.2 \bigstrut[b]\\
    \hline
    250 & nearest node & 1,579.6$\pm$65.4 & 1,674.6$\pm$83.2 & 1,762.0$\pm$102.7 & 1,864.9$\pm$97.7 & 1,951.1$\pm$97.7 & 2,055.0$\pm$120.1 \bigstrut[t]\\
    250 & closest vehicle & 1,578.0$\pm$57.9 & 1,669.5$\pm$87.7 & 1,759.6$\pm$112.0 & 1,858.7$\pm$83.8 & 1,957.8$\pm$95.4 & 2,062.7$\pm$126.0 \bigstrut[b]\\
    \hline
    300 & nearest node & 1,687.9$\pm$72.5 & 1,773.4$\pm$66.7 & 1,870.7$\pm$93.4 & 1,958.7$\pm$104.2 & 2,059.1$\pm$117.2 & 2,159.7$\pm$134.2 \bigstrut[t]\\
    300 & closest vehicle & 1,695.7$\pm$67.2 & 1,770.6$\pm$74.1 & 1,893.4$\pm$72.0 & 1,983.1$\pm$110.9 & 2,076.7$\pm$112.1 & 2,184.3$\pm$119.3 \bigstrut[b]\\
    \hline
    350 & nearest node & 1,809.3$\pm$61.0 & 1,887.1$\pm$73.9 & 1,969.7$\pm$75.9 & 2,060.8$\pm$92.6 & 2,133.7$\pm$95.8 & 2,232.4$\pm$125.0 \bigstrut[t]\\
    350 & closest vehicle & 1,824.7$\pm$68.2 & 1,903.4$\pm$66.0 & 2,003.7$\pm$72.1 & 2,078.1$\pm$80.1 & 2,167.9$\pm$99.9 & 2,260.7$\pm$117.4 \bigstrut[b]\\
    \hline
    400 & nearest node & 1,927.1$\pm$63.3 & 2,024.7$\pm$76.7 & 2,092.4$\pm$76.3 & 2,186.2$\pm$81.6 & 2,280.0$\pm$121.9 & 2,366.5$\pm$117.3 \bigstrut[t]\\
    400 & closest vehicle & 1,960.0$\pm$79.4 & 2,032.5$\pm$66.4 & 2,109.5$\pm$80.8 & 2,190.3$\pm$76.2 & 2,294.3$\pm$99.2 & 2,394.3$\pm$107.3 \bigstrut[b]\\
    \hline
    450 & nearest node & 2,035.2$\pm$79.3 & 2,105.1$\pm$95.5 & 2,198.5$\pm$94.4 & 2,297.3$\pm$103.1 & 2,370.6$\pm$96.5 & 2,477.2$\pm$128.9 \bigstrut[t]\\
    450 & closest vehicle & 2,038.4$\pm$78.7 & 2,101.3$\pm$96.0 & 2,213.8$\pm$104.1 & 2,298.5$\pm$102.2 & 2,382.9$\pm$103.8 & 2,479.9$\pm$129.9 \bigstrut[b]\\
    \hline
    500 & nearest node & 2,128.9$\pm$62.7 & 2,211.5$\pm$68.9 & 2,290.5$\pm$100.0 & 2,380.6$\pm$110.1 & 2,455.3$\pm$116.2 & 2,528.8$\pm$109.2 \bigstrut[t]\\
    500 & closest vehicle & 2,127.8$\pm$58.1 & 2,197.8$\pm$78.4 & 2,326.4$\pm$90.9 & 2,390.5$\pm$105.9 & 2,482.2$\pm$97.4 & 2,549.3$\pm$139.7 \bigstrut[b]\\
    \hline
    \end{tabular}%
}
  \label{tab:heuristics-medium-instances}%
\end{table*}%
Figure \ref{fig:dist-relation} (a) shows the average solution distance growth
when the number of salesmen (trucks) increases for the 50 nodes test-instance for the closest vehicle heuristic.
This represents row 2 in Table \ref{tab:heuristics-medium-instances}.
The growth of the average solution distance with the number of nodes is shown in Figure \ref{fig:dist-relation} (b).
The $m=2$ column for the closest vehicle heuristic in Table \ref{tab:heuristics-medium-instances} shows the figure's averages 
and additionally standard deviations are given.

\begin{figure} \centering
\includegraphics[height = 5cm, width=\columnwidth, keepaspectratio]{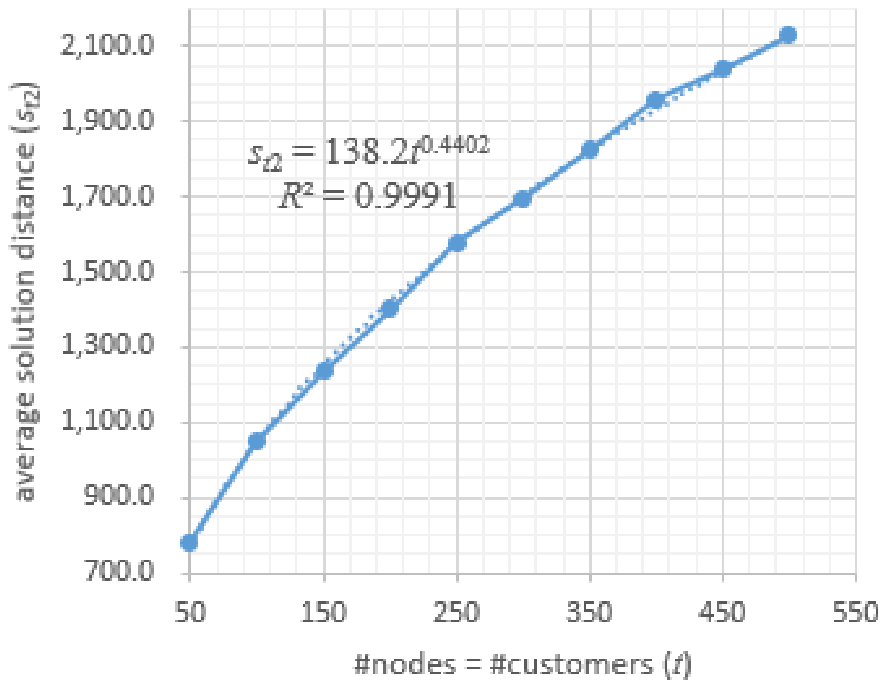}%
\caption{Average total distance dependent on number of nodes for the closest vehicle heuristic with $m=2$.}%
\label{fig:dist-nodes-growth}%
\end{figure}

The above allows to estimate the solution distance for the closest vehicle heuristic
depending on the number of travelling salesmen.
We observe solution distance changes between $m$ and $m+1$ (salesman between 2 and 7)
for the test-instance (50 to 500).
This is on average $\Delta = 90.40$ (with standard deviation 16.5) for the closest vehicle heuristic,
and $\Delta = 88.09$ (with standard deviation 13.0) for the nearest neighbour algorithm.
These values were obtained from:
\lequ{avg-delta}{ \Delta = \frac{1}{n_s n_m} \sum_{t \in S} \sum_{m=3}^{\max m} s_{tm} - s_{t,m-1}, }
where $S = \set{50, 100,\dots, 500}$, $n_s = \#S$ and $n_m = \max m -2 = 5$. 
Hence, the solution distance $s_{tm}$ can be approximated by:
\lequ{distance-m}{ \tilde{s}_{tm} =  s_{t2} + (m-2) \Delta.}
$s_{t2}$ itself can be expressed by fitting a function $x^p$ to the data.
This gives the following functional relationship for the closest vehicle heuristic for two salesmen:
\lequ{dist-nodes}{s_{t2} \sim 138.2 t^{0.44},}
using the previous data (see Table \ref{tab:heuristics-medium-instances} and Figure \ref{fig:dist-nodes-growth}).
For instance, $t = 250$ and $m=2$ returns $s_{250,2} = 1,568.9$ (actual average 1578.0). 
Combining Equation (\ref{eq:distance-m}) and (\ref{eq:dist-nodes}) gives the functional relationship:
\lequ{final-mTSP}{\tilde{s}_{tm} \approx   138.2 t^{0.44}  + 90.4(m-2),~m \geq 2.}
The above describes the total distance covered by a fleet of $m$ vehicles servicing $n_s$ customers.
For instance, for $m=6$ vehicles and $t=250$ customers Equation (\ref{eq:final-mTSP}) 
returns a travelled distance of $s_{250,6} = 1930.5$ (actual average 1957.8, 1.392\% deviation) in $X^2$.
This can be easily scaled to larger areas. For instance, if $X^2=300\times300\text{km}^2$ then $s_{tm}$ is multiplied by three (slightly more precise $3.02=\frac{299}{99}$).
In the special case of $m=1$ the nearest neighbour and closest vehicle heuristic return the same solution for the TSP.
The expected distance travelled $s_{t1}$ using either heuristic is: $s_{t1} =111.37t^{0.4704}$, where $t$ is the number of customers.

Inspired by this empirical information 
an even more mathematical approach is considered.
Fundamental probability theory provides 
the expected distance between two uniformly distributed random numbers $a(X)$.
Here, $X$ is a compact convex subset of the $s$-dimensional Euclidean space.
Let $d(X)=\max \{\|x-y\| : x, y \in X\}$ be the diameter of $X$.
It is well known that for all compact convex subsets of $\mathbb{R}$ 
the expected distance is $a(X)=\frac{d(X)}{3}$.
\begin{example}[Expected distances between two points]
Let $a=0$ and $b=100$. The probability for each number is $\frac{1}{b-a}$.
The expected value for two nodes $n=2$ is $\frac{b-a}{3}=\frac{100}{3}=33.\dot{3}$.
For three numbers the problem becomes more interesting.
\end{example}

\begin{example}[Numerical simulation - average distance between two points]
An easy way to confirm the average distance of two points is to run a numerical simulation.
I have created $n=100,000$ points with $x, y \in X^n; X = U(0,1))$. 
The distance vector is $d=|x-y|$ with $d(X)=1$.
The average distance is $a(X) = \frac{1}{n} e d \sim 0.33321$, where $e$ is the one-vector.
The distribution is triangular and shown in Figure \ref{fig:sim-avg-dist-2-points} (a).
\begin{figure*}[htbp]\centering
	\begin{tabular}{ccc}
		\includegraphics[width=.4\textwidth,height=4cm, keepaspectratio]{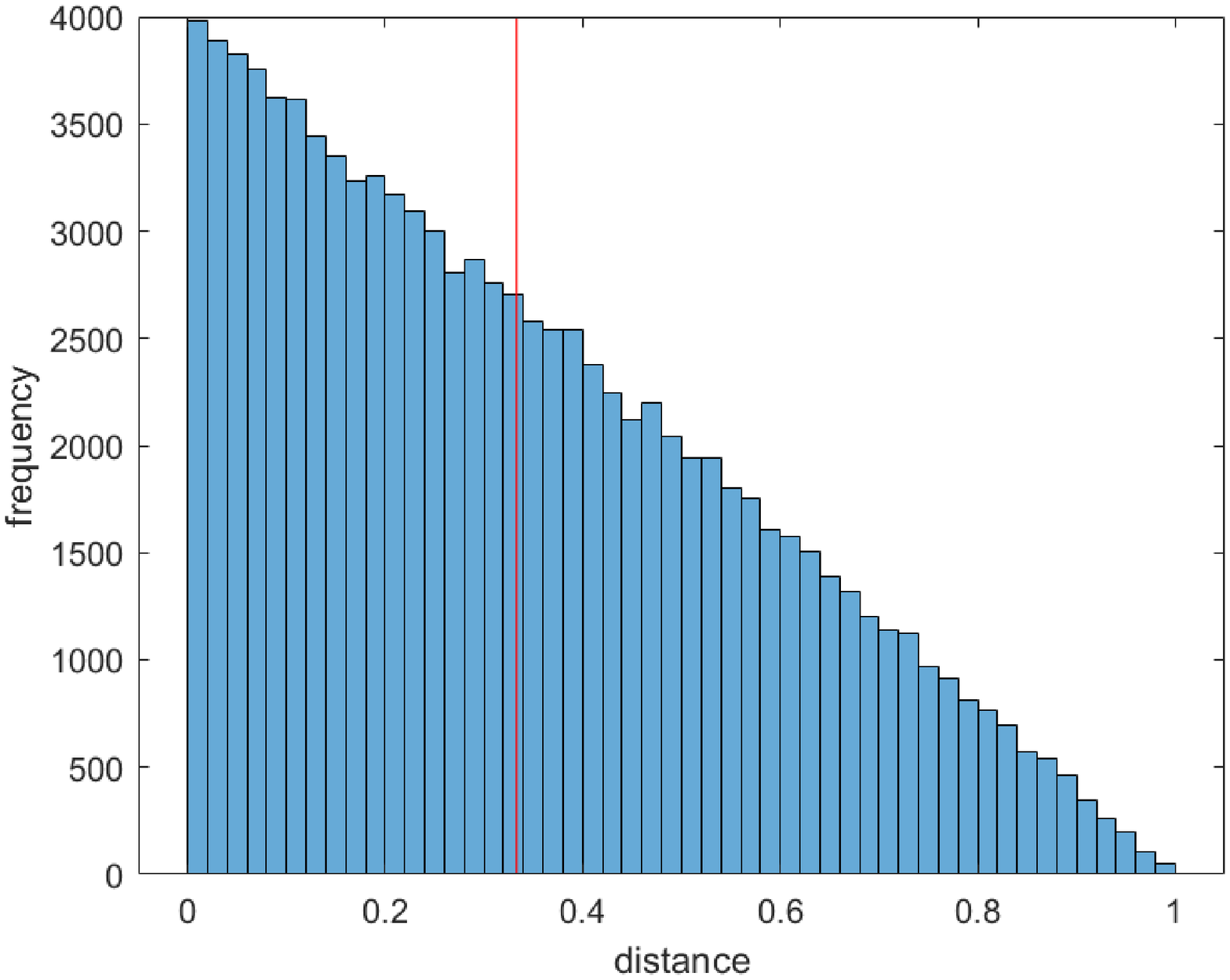}&
	  \includegraphics[width=.4\textwidth,height=4cm, keepaspectratio]{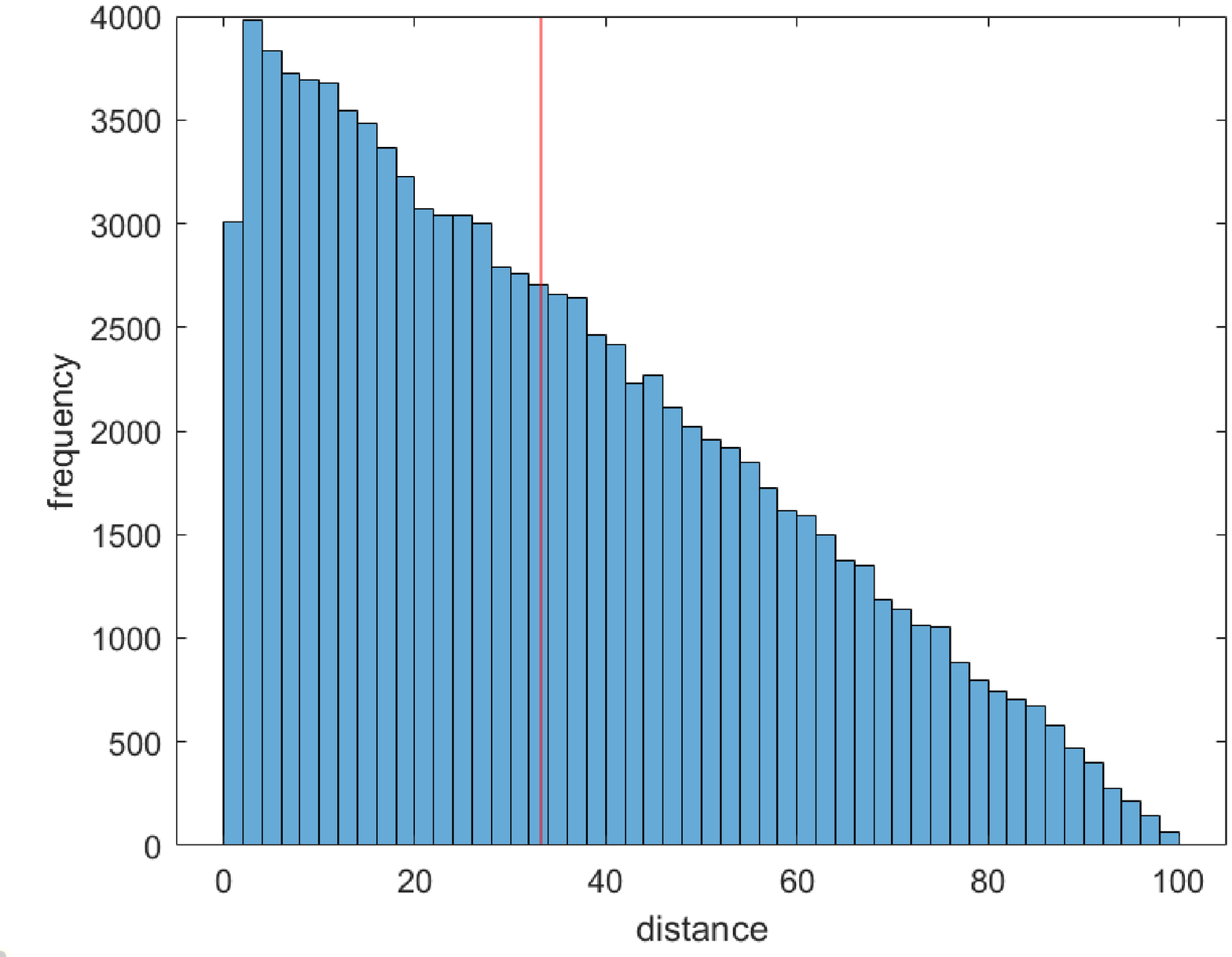}\\
		(a) & (b)
	\end{tabular}
	\caption{Numerical simulation - frequency distribution of distances of two points 
	(a) U(0,1); (b) integers in \sett{100}.}%
	\label{fig:sim-avg-dist-2-points}%
\end{figure*}
This agrees very well with the theoretical average $a(X)=\frac{d(X)}{3}$.

Let us look at 100,000 uniformly distributed integers in $X=\sett{100}$.
The observed average distance between them is $a(X) = 33.1668$ and $d(X) = 99$.
The distance distribution is shown in Figure \ref{fig:sim-avg-dist-2-points} (b).
Note, that the assumptions for the above formula are not met, but $a(X) \sim \frac{1}{3} d(X)$.
\end{example}

\begin{example}[Numerical simulation - minimum distance to $n$ points]
To get an idea about the minimum distance to $n$ points a numerical simulation with $m=10^5$ repetitions is done.
The minimum mean distance from $x_k \in U(0,1)$ to $y_k \in U(0,1)^n, n=50$ points 
is $a(U^n) = \frac{1}{m} \sum_{k=1}^{m} \min|e\cdot x_k-y_k| = 0.009984719 \approx 0.01$, where $e$ is the one vector.
Figure \ref{fig:sim-min-dist} (a) shows the distribution of the minimum mean distances.
\begin{figure*}[htbp]\centering
	\begin{tabular}{ccc}
		\includegraphics[width=.4\textwidth,height=4cm, keepaspectratio]{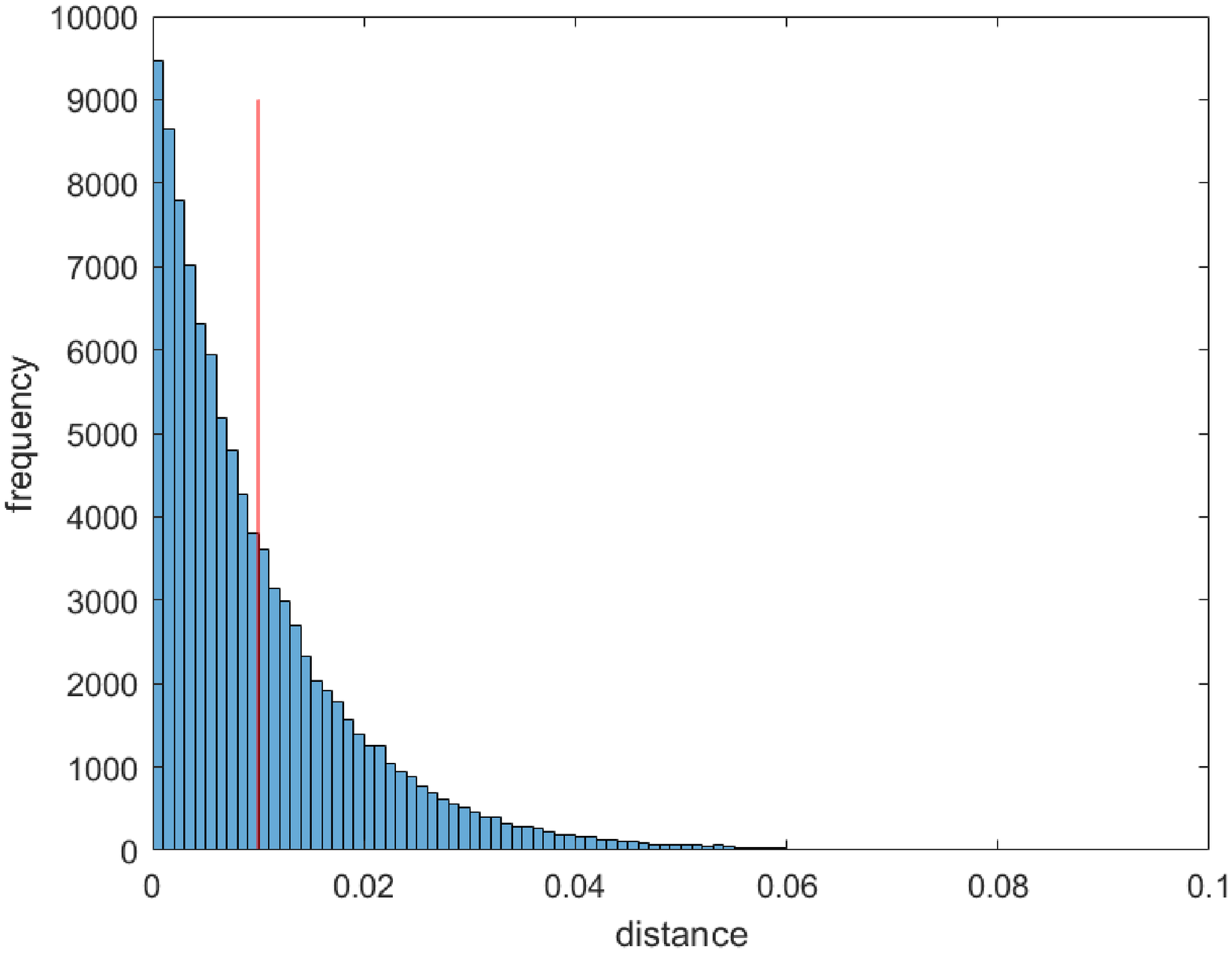}&
	  \includegraphics[width=.4\textwidth,height=4cm, keepaspectratio]{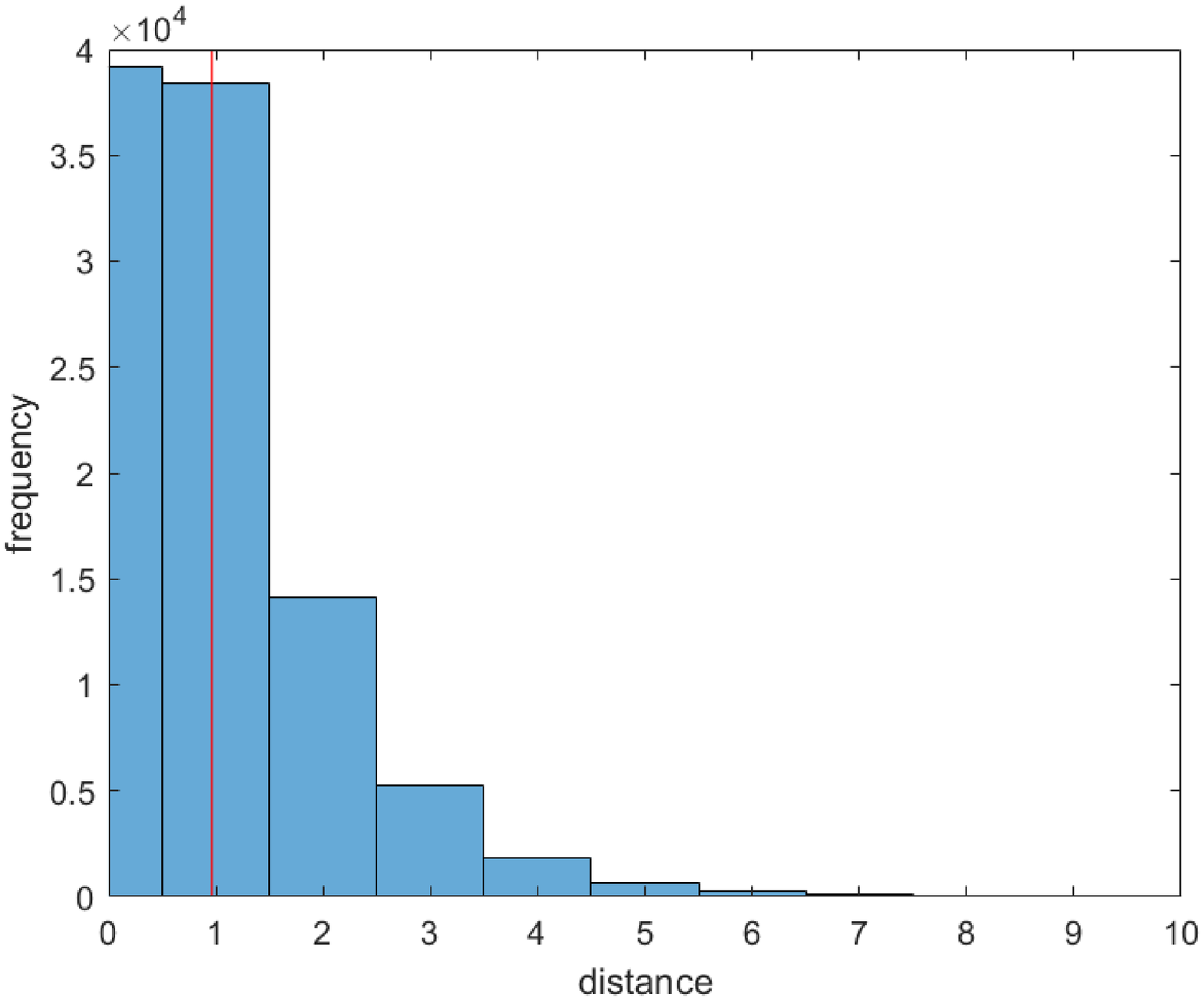}\\
		(a) & (b)
	\end{tabular}
	\caption{Numerical simulation - frequency distribution of minimum distances to 50 uniformly distributed points 
	(a) $U(0,1)$ to $U(0,1)^n$; (b) similarly but integers in \sett{100}.}%
	\label{fig:sim-min-dist}%
\end{figure*}
	It can be seen that $\min|e\cdot x_k-y_k|$ distribution has a negative exponent.
	
	In the integer domain $\sett{100}$ a similar experiment ($n=50, m=10^5$) is done.
	The observed average distance between them is $\bar{d}\approx0.9644$.
	The distance distribution is shown in Figure \ref{fig:sim-min-dist} (b).
	We see that distance zero and one are the most frequent ones.
	The zero distances are due to samples $y$ including $x$, hence, $a(\sett{100}^n)<1$.
\end{example}

Considering the minimum distance 
from a uniformly distributed point $x \in U(0,1)$
to $n$ uniformly distributed points $y \in U(0,1)^n$:
\lequ{min-dist-U01}{ m_d = \min |e x - y|, }
where $e$ is the one vector.
\begin{figure}[htbp]\centering
\includegraphics[width=\columnwidth,height=6cm, keepaspectratio]{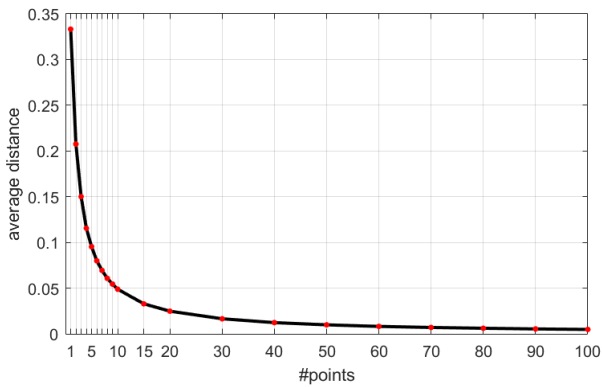}%
\caption{Minimum average distance from a point $x$ to $n$ points in $U(0,1)$.}%
\label{sim-min-dist-U01}%
\end{figure}
Generally, the following equation: 
\lequ{min-dist-U01n}{a(U^n) = 0.4158n^{-0.949}}
describes the relationship well for $n>1$.
This is an important finding, since one of the main steps in both heuristics is to determine the minimum distance.
The nearest node Algorithm \ref{alg:mTSP-nearest-node} finds the minimum in line 5;
the closest vehicle Algorithm \ref{alg:mTSP-closest-vehicle}  obtains it in line 6.

In the $s$-dimensional case \citet{sors2004integral} have documented the following finding.
If $X \subseteq \mathbb{R}^{s}$ is a ball with diameter $d(X)$, then
\equ{a(X)=\frac{s}{2 s+1} \beta_{s} d(X),}
where
\equ{
\beta_{S}=\left\{\begin{array}{ll}{\frac{2^{3 s+1}((s / 2) !)^{2} s !}{(s+1)(2 s) ! \pi}} & {\text { for even } s} \\ {\frac{2^{s+1}(s !)^{3}}{(s+1)(((s-1) / 2) !)^{2}(2 s) !}} & {\text { for odd } s}\end{array}\right.}
The two-dimensional case simplifies to:
$a(X)=64 d(X) /(45 \pi)$.

Our experimental design was based on rectangles. 
\citet{dunbar1997average} have derived the following:
\equ{\begin{aligned} 
a(X)=\frac{1}{15}[ & \frac{a^{3}}{b^{2}}+\frac{b^{3}}{a^{2}}+d\left(3-\frac{a^{2}}{b^{2}}-\frac{b^{2}}{a^{2}}\right) \\ 
&+\frac{5}{2}\left(\frac{b^{2}}{a} \log \frac{a+d}{b}+\frac{a^{2}}{b} \log \frac{b+d}{a}\right) ],
\end{aligned},}
where $d=d(X)=\sqrt{a^{2}+b^{2}} .$ 

If $X$ is a square, then the above simplifies to:
\lequ{avg-dist-2-points}{ \begin{aligned}
	a(X)&=(2+\sqrt{2}+5 \log (\sqrt{2}+1)) \frac{d(X)}{15 \sqrt{2}}\\
	    &\sim 0.36869 d(X) \end{aligned}}

\begin{example}[Average distance of two points]
Assume a square with length 100. Hence, $d(X) = \sqrt{2} \cdot 100$.
Equation (\ref{eq:avg-dist-2-points}) allows us to compute 
the average distance of two points in the Euclidean plane.
This gives us: $a(X) \sim 0.36869 d(X) =  0.36869 \sqrt{2} \cdot 100 =\sim 52.14$.
\end{example}

\citet{burgstaller2009average} have found more general bounds on the average distance 
between two uniformly distributed points which are independently chosen from a compact convex subset $X$
in an $s$-dimensional Euclidean space. 
This problem is formulated as:
\lequ{prob-dist}{a(X) =\mathbb{E}[\|x-y\|]=\frac{1}{\lambda(X)^{2}} \int_{X} \int_{X}\|x-y\| d \lambda(x) d \lambda(y)}
where $x$ and $y$ are two points and $\lambda$ denotes the $s$-dimensional Lebesgue measure.

The depot is located between $a$ and $b$ being a uniformly distributed random integer $c$. 
We begin our considerations with single salesman and the closest neighbour algorithm 
and the coordinates of customers being integers on a line between $a$ and $b$.
If there are sufficiently many customers, the uniform distribution means that customers are equally likely to be located at the borders.
The algorithm first visits customers on one side covering $d_1 = c-a$.
Followed by crossing over the depot ($d_2 = c-a+1$) and visiting the customers on the other side $d_3=b-c$.
Finally returning to the depot $d_4=b-c$.
This means the expected distance is: \equ{2(b-a)+1.}

\begin{figure*}\centering
\begin{tabular}{cc}
\includegraphics[width=.45\textwidth, height=6cm, keepaspectratio]{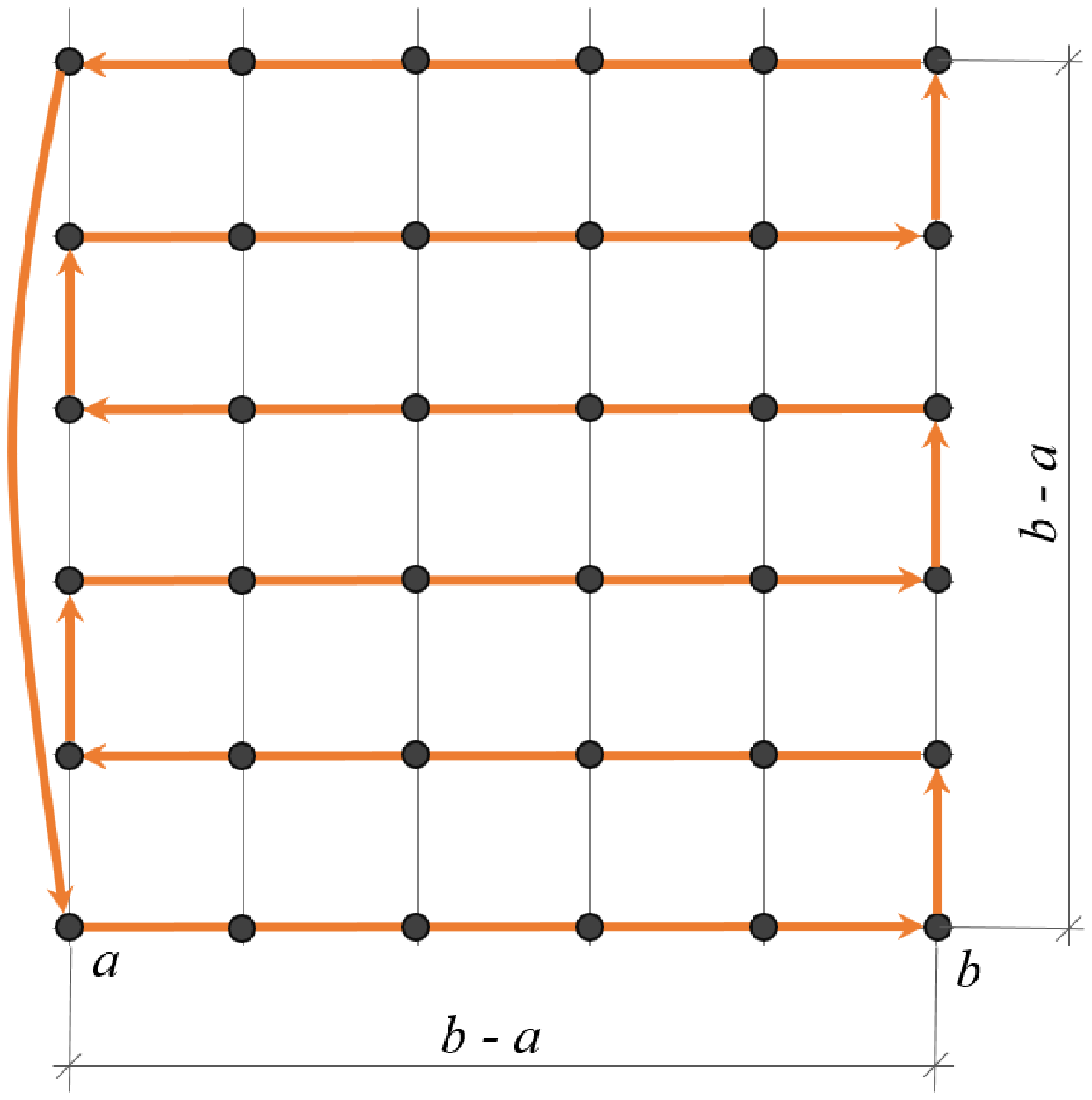}&
\includegraphics[width=.45\textwidth, height=6cm, keepaspectratio]{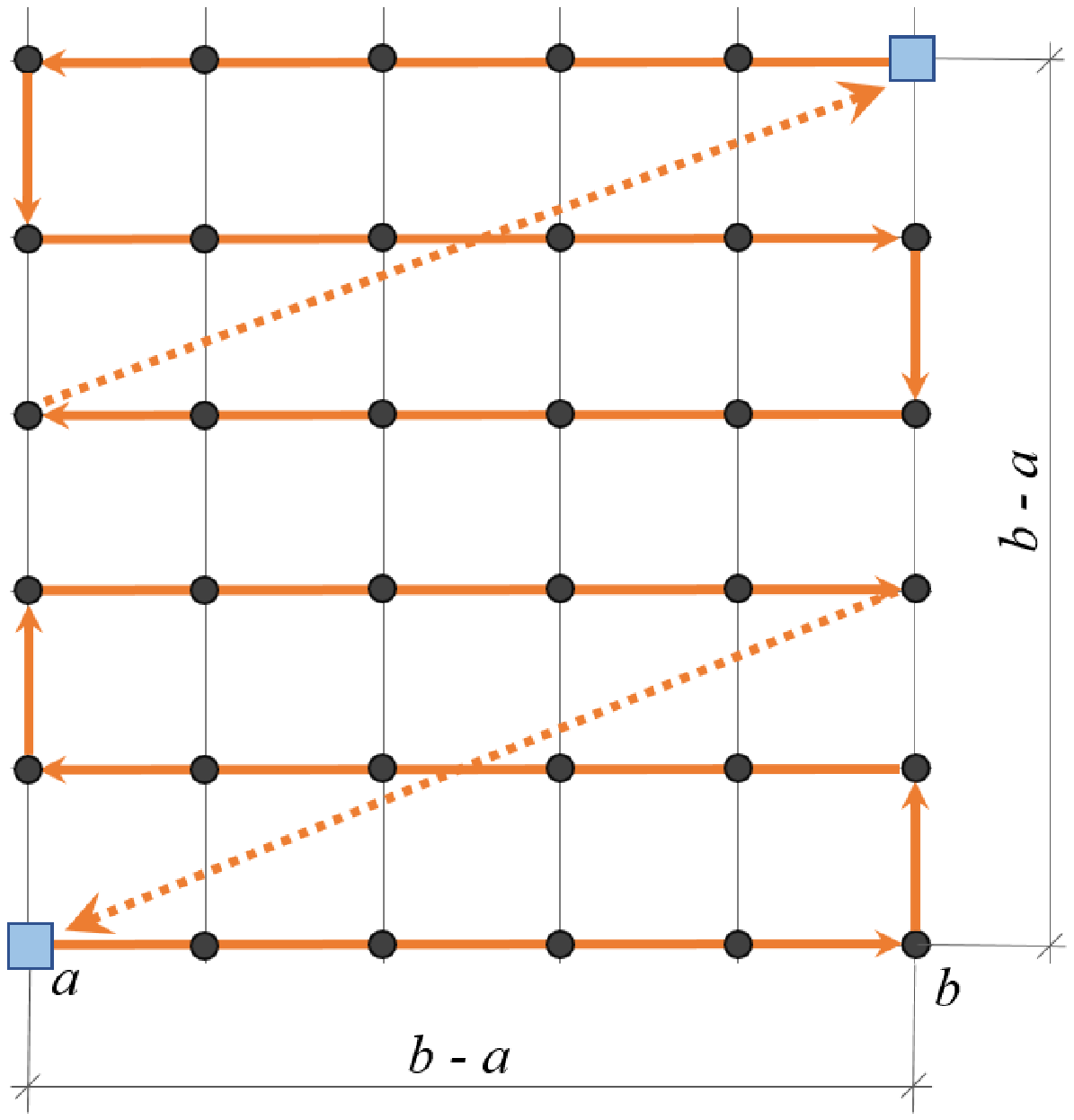}\\
(a) & (b)
\end{tabular}
\caption{Nearest node (a) TSP with single depot; (b) mTSP with two opposite corner depots.}%
\label{fig:TSP-greedy}%
\end{figure*}

Let us now consider the two-dimensional case covering the area $(b-a)^2$.
Assume that customers are located at each grid point.
We modify the closet neighbourhood algorithm such 
that going sideways is preferred when faced with similar distances.
Otherwise the algorithm ends up in a kind of random walk causing many unnecessary ``jumps''.
Figure \ref{fig:TSP-greedy} (a) gives an illustration of the traversal for one salesman, i.e. a TSP - starting at $(a,a)$.
Hence, the travelled route length is: \equ{(b-a)(b-a+1) + 2(b-a) = (b-a)^2 + 3(b-a).}

The travelled distance for two salesmen with balanced workload, fixed opposite corner depots
and $b-a$ being an even number -
as illustrated in Figure \ref{fig:TSP-greedy} (b) is:
\lequ{2TSP-corner}{\begin{aligned}
(b-a)(b-a+1)+b-a-1+\\ 2 \sqrt{(b-a)^{2}+\frac{1}{4}(b-a-1)^{2}} = \\
(b-a)^2+2(b-a)-1+\\ \sqrt{5 a^{2}-10 a b+2 a+5 b^{2}-2 b+1}
\end{aligned}
}
A general formula for the single corner depot case with two salesmen can be derived using a similar approach.

\begin{example}[2TSP]
Let $a=1$ and $b=100$. 
Assume a customer is in each grid point, i.e. a total of $100^2 = 10,000$.
There are two salesmen.
(a) The depots are located in opposite corners.
Equation (\ref{eq:2TSP-corner}) gives the distance:
\equ{99^2+2\cdot 99-1+2\sqrt{99^2 + \frac{1}{4}98^2} \sim 10,110.5.}
\end{example}

Future work can take this as starting point 
to derive the exact relationship between distance and any number of salesmen.

\section{Conclusion}\label{sec:Conclusion}
The exact mTSP with balancing constraints is the preferable solution 
for small instances, e.g. \#nodes less than $n=100$ in a complete network.
For larger instances runtime becomes an issue for exact methods,
yet heuristics find solutions in $O(n)$.
The closest vehicle and nearest node heuristics were introduced here.
They have similar solution values and runtimes.
The closest vehicle heuristic can be used when bargaining for jobs in a real-world scenario.
For instance, a truck has done a delivery and is assigned the closest customer as next drop-off location.
The nearest neighbour heuristics can be used in a real-world collection scenario, 
such as a single truck doing all pick-ups but has limited capacity.
Comparing these heuristics to meta-heuristics from other studies (using the same test-instances)
show surprisingly similar solution distances.

The second contribution is a closed form formula that describes the total mTSP-distance
dependent on the number of vehicles and customers.
Adding more salesmen to a scenario increases the total distance, 
because of the additional leaving and entering arc.
Trivially, each vehicle's distance reduces; implying customers can serviced within a shorter time window.
As starting point for the theoretical underpinnings, the expected distance of two uniformly distributed random points (real and integer) was reviewed. 
Additionally, the average minimum distance from one point to $n$ points in $U(0,1)$ and $\sett{n}$ were analysed,
since this is at the core of the proposed mTSP heuristics. 
The relationship was expressed in Equation (\ref{eq:min-dist-U01n}). 
Further work is required to derive the precise statistical distance growth formula for the mTSP.
However, a simulation of uniformly distributed nodes in a Euclidean grid plane shows, 
when the number of salesman is increased the distance grows approximately linearly.
Furthermore, the growth in distance is roughly proportional to the square root of \#customers.
Combining those two relationships gives the closed form formula (\ref{eq:final-mTSP}) in $\sett{n} \times \sett{n}$,
which expresses the expected mTSP distances the heuristics would have found.
It is believed that a similar relationship holds for the exact methods and other heuristics in the Euclidean plane.
Hence, giving an efficient way to estimate the distance increase and time reduction when vehicles are added or customer numbers change.

\section{Acknowledgement} 
I would like to thank the inspiring and creative Hyster-Yale UK team, 
who's work on the next generation of intelligent autonomous forklift-trucks motivated me to look for mTSP solutions. 
Mark's vision of forklift-trucks bargaining for the next job inspired the closest vehicle mTSP heuristic. 
Chi (Ezeh) encouraged me to keep a wholistic view that reaches from strategic mTSP formulations to vehicles' Artificial Intelligence. 
Chris drew my attention to concurrency issues, that vehicles encounter when using the same path. 
There were many more inspirations, which demonstrated the importance of the mTSP and its solution approaches in practical applications. 
\bibliography{mTSP}

\end{document}